\documentclass[runningheads]{llncs}
\usepackage{graphicx}
\usepackage{hyperref}
\usepackage{makeidx} 
\usepackage[utf8]{inputenc}
\usepackage{amsmath}
\usepackage{graphicx}

\begin{document}

%
\title {Self-supervised Recurrent Neural Network for 4D Abdominal and In-utero MR Imaging}
\titlerunning{4D Abdominal and In-utero MRI}
\author{Tong Zhang\inst{1}\thanks{tong.zhang@kcl.ac.uk}, Laurence H. Jackson\inst{1}, Alena Uus\inst{1},  James R. Clough\inst{1}, \\ Lisa Story\inst{2}, Mary A. Rutherford\inst{1}, Joseph V. Hajnal\inst{1}, and Maria Deprez\inst{1}}
\institute{School of Biomedical Engineering and Imaging Sciences, King's College London, London, United Kingdom\\
\and Department of Women and Children’s Health,  School of Life Course Sciences, King's College London, London, United Kingdom}
\authorrunning{T. Zhang et al.}

%
\maketitle              
\begin{abstract} Accurately estimating and correcting the motion artifacts are crucial for 3D image reconstruction of the abdominal and in-utero magnetic resonance imaging  (MRI). The state-of-art methods are based on slice-to-volume registration (SVR) where multiple 2D image stacks are  acquired in three orthogonal orientations.
In this work, we present a novel reconstruction pipeline that only needs one orientation of 2D MRI scans and can reconstruct the full high-resolution image without masking or registration steps. 
The framework consists of two main stages: the respiratory motion estimation using a self-supervised recurrent neural network, which learns the respiratory signals that are naturally embedded in the asymmetry relationship of the neighborhood slices and cluster them according to a respiratory state.
Then, we train a 3D deconvolutional network for super-resolution (SR) reconstruction of the sparsely selected 2D images using integrated reconstruction and total variation loss. We evaluate the classification accuracy on 5 simulated images and compare our results with the SVR method in adult abdominal and in-utero MRI scans. 
The results show that the proposed pipeline can accurately estimate the respiratory state and reconstruct 4D SR volumes with better or similar performance to the 3D SVR pipeline with less than 20\% sparsely selected slices. The method has great potential to transform the 4D abdominal and in-utero MRI in clinical practice.
\end{abstract}

\section{Introduction}
Due to its non-invasive nature and the superior soft-tissue contrast, magnetic resonance imaging (MRI) is becoming increasingly more popular for the adjunct pregnancy screening~\cite{Lloyd2019,story2017magnetic,STORY2018134,story2019magnetic}. The typical scanning time for a 2D MR stack that covers the whole fetus and placenta varies from 1 to 10 minutes per stack depending on the MR sequences, field-of-view (FoV) and slice thickness. The long scanning time inevitably introduces a series of motion artifacts, such as maternal breathing, organ deformation, and fetal movement.

\textbf{In utero MR image reconstruction}. To reconstruct a high quality 3D or 4D placenta and in-utero MRI, accurately estimating the respiratory motion is a key step. The current state-of-the-art methods correct the through-plane motion using slice-to-volume registration (SVR) reconstruction pipelines~\cite{Gholipour2010,Kuklisova-Murgasova2012,Kainz2015,Ebner2018,TORRENTSBARRENA2019}. These methods require three to nine 2D image stacks to be acquired in orthogonal orientations; then a region-of-interest (ROI) mask needs to be generated manually or automatically~\cite{Ebner2018,TORRENTSBARRENA2019}; finally the 3D SR image is  iteratively generated based on the optimisation of the SVR results, robust statistics, intensity correction, and estimated point spread function (PSF). The acquisition of multiple 2D MR stacks in different orientations is time-consuming compared to the single orientation and inevitably introduces motion artifacts. The registration methods in the SVR pipelines are rigid and cannot correct the non-rigid respiratory motion. The SVR pipelines thus need enough redundant 2D images to reject all the slices where deformation compared to the reconstructed volume occurred. The overall image reconstruction performance will depend on the accuracy of the ROI masking, registration and data redundancy.

\textbf{Self-supervised learning} is generally considered as a subset of unsupervised learning, where the extensive cost of manual annotations is avoided and replaced by supervisory signals or automatically generated labels. Compared to the popular supervised methods which train the neural network with paired data $Xi$ and label $Yi$, the self-supervised methods train with data $Xi$ 
with its pseudo label $Pi$, which
is generated automatically without involving any human annotation. Several recent papers have explored the usage of the temporal ordering of frames/images as a supervisory signal for complex video analysis~\cite{ramanathan2015learning,fernando2017self,Wei2018}. In particular, Wei et al.~\cite{Wei2018} explored detecting and learning the direction of time for action recognition and video forensics. 

\textbf{Contribution} In this work, we propose a respiratory motion resolved 4D (3D+t) reconstruction pipeline of a single orientation stack of 2D MR slices, based on an bidirectional self-supervised recurrent neural network (RNN)~\cite{graves2005framewise} for identification of breathing states and efficient modified balanced steady state free precession (bSSFP) sequence with the SWEEP technique~\cite{LaurenceHJacksonAnthonyNPriceJanaHutterLucilioCordero-GrandeAlisonHoPaddyJSlator2018}. The method does not require masking or registration. Our experimental results show that the SWEEP MR acquisition in combination with the proposed pipeline enables 4D (3D +t) SR reconstruction of abdominal and in-utero images, and outperforms the SVR for 3D reconstruction with using less than 20\% total slices for each respiratory state. To the best of our knowledge, it is the first successful application of self-supervised network for image-driven respiratory motion estimation and 4D(3D+t) MR SR reconstruction in the medical imaging community.

\section{Method}

\subsection{Data acquisition}

A stack of MR slices is acquired sequentially using a modified bSSFP SWEEP sequence~\cite{LaurenceHJacksonAnthonyNPriceJanaHutterLucilioCordero-GrandeAlisonHoPaddyJSlator2018} which allows fast acquisition of large number of densely spaced overlapping slices, thus providing sufficient information for local estimation of respiratory motion. SWEEP continuously shifts the radiofrequency excitation frequency so as to maintain a single stable signal state across a volume, negating the requirement for start-up cycles and resulting in a maximally efficient acquisition for dense slice sampling applications. The acquisition time per slice is 490ms for the uterus scans and 442ms for the kidney scans, which freezes nearly all in-plane respiratory motion. The total scan time depends on the total slice number which is 3 to 10 minutes. This sequence also minimises the effects of fetal motion, by minimising the time between acquisition of the neighbouring slices while maintaining high MR signal. This effectively removes the need for masking, as the data is locally consistent except for the respiratory motion.

\subsection{The reconstruction pipeline}

The reconstruction pipeline consists of cascading a self-supervised RNN to estimate the respiratory states for each slice and a three layer super-resolution (SR) neural network (SR-net) for reconstruction respiratory-state specific 3D volumes using the respiratory state classes predicted by RNN. The overall pipeline is summarised in Fig.$1$.
\begin{figure}
\includegraphics[width=1\textwidth]{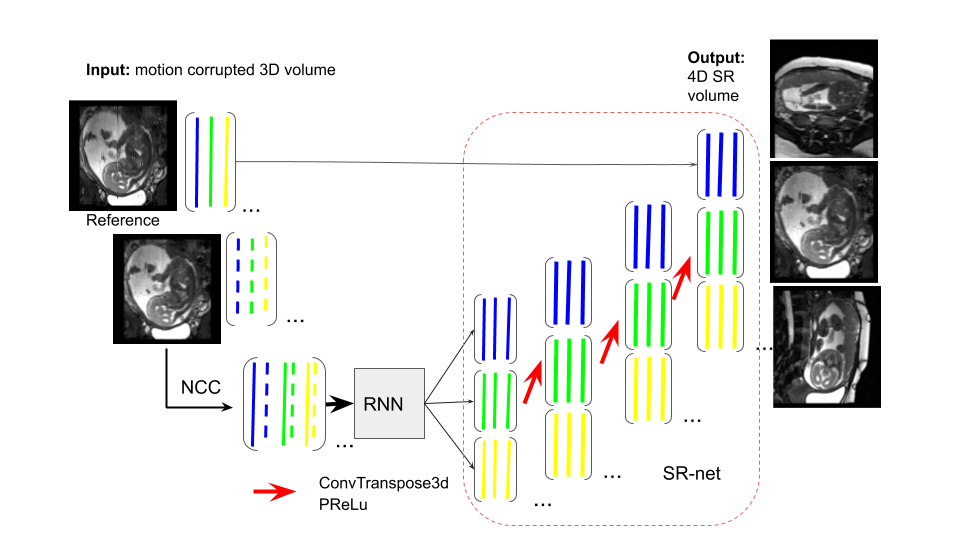}
\caption{Reconstruction pipeline for respiratory motion resolved 3D+t abdominal and in-utero MRI.}
\label{fig:zero}
\end{figure} 
%
\subsection{Self-supervised RNN}
Due to sequential acquisition of slices when using the SWEEP sequence, the respiratory signal is embedded in the neighborhood slices in the arrow of acquisition time. To separate the slices into different respiratory states, we train a bi-directional self-supervised RNN (SRNN). 

We first generate a reference volume based on 1D convolution with a Gaussion kernel along the acquisition axis (Z-axis). Intuitively, the reference volume is most similar to the average states of inhale and exhale. We then calculate the normalised cross-correlation (NCC) between each slice in the motion corrupted volume and the reference volume and the average inhale and exhale states are identified as the peaks of NCC sequence. We then separate those two average states based on their timing orders. The remaining states are identified linearly based on the distance between the average states. The approximate states automatically determined by this approached are then used to train bi-directional RNN.

Consider the input motion corrupted MRI scan as a group of 2D image sequence $ I= \left\{ I_{1},I_{2},...,I_{T} \right\} $ , where the slice number is equivalent to the arrow of time. For analysing temporal features, we use the bidirectional LSTM network to formulate the respiratory states that naturally embedded in the neighbourhood slices, where both past and future events is used for prediction~\cite{graves2005framewise}. Each LSTM unit computes the hidden vector sequence $ h =  \left\{h_{1},h_{2}...,h_{T} \right\}$ and  memory cell $ C=  \left\{C_{1},C_{2}...,C_{T} \right\}$ and output vector sequence $y = \left\{y_{1},y_{2},...y_{T}\right\}$ by bidirectional iterating from the sequence time $ t=1$ to $T$ and $ t=T$ to $1$. We built a three layer bidirectional LSTM and set the total classes to 10 in the fully connected layer. In this work, we automated annotate each slice with a respiratory state, then segment the volume into multiple 20-slice subvolumes with 1 slice overlap, the input of SRNN is a $20\times 20$ cosine similarity matrix and the output is the last slice prediction. 

\subsection{Super-resolution reconstruction (SR-net)}
Deep learning based SR methods which trained on paired low resolution and SR images are reported to outperform the traditional ones~\cite{Dong2016Image}. Our method offer the first time non-example based SR solution, which use PSF as downgrade function and jointly penalize the MSE and TV losses. 
For each respiratory state, the selected slices are used to perform SR reconstruction. As shown in Fig.1, we train a four layer 3D ConvNet with parametric rectified linear unit (PReLu), where the loss function is defined as the combined reconstruction error and the total variation (TV) regularisation~\cite{doi:10.1137/090769521}. As proposed previously~\cite{Kuklisova-Murgasova2012}, we treat PSF as a 3D Gaussian function with Full width at half maximum (FWHM) equal to the slice-thickness in the through-plane direction. The reconstruction error then can be expressed as $E(V)=\sum_{jk}(R_{jk}-S_{jk})^2$ where $R_{jk}$ refers to the intensity of the voxel $j$ in each selected slice indexed by $k$ and $S_{jk}$ are that simulated from isotropic super-resolved volume $V$ using the PSF. The loss function of SR-net is formulated as:
\begin{equation}
L_{SR} = E(V) + \sum\limits_{i} \lambda_{i}\sum\limits_{l} TV^{1D}(V(d_{i})_{l})
\end{equation}

where $ \lambda_{i} $ is weighting coefficient that balances the TV loss in different orientations, and $V(d_{i})$ denotes every possible $l$-dimensional slice of $V$ following dimension $d_{i}$. SRnet takes less than 1 min at test time for a 4D reconstruction of our data, while previously proposed methods that were build to handle randomly oriented PSFs take around 40 mins [5] and 5 hours [6]. 

\subsection{Implementation Details}
The method was implemented using Python and Pytorch. The network was trained in two steps, first the SRNN is trained with 1359 subvolumes from 2 subjects with 8 groups of breathing states. Then, the SRnet is trained on 24 3D volumes from 3 subjects. For both SRNN and SRnet training, we use Adam as an optimisation tool. For SRNN, the learning rate has been tested from 0.01 to 1 and set to 0.1 based on empirical results. To avoid the over-fitting, we set the weight decay to 0.01, which add L2 regularization of the  weights into the optimisation procedure. For SR-net, we set the TV loss weights to $0.01$, $0.01$, and $0.1$ to enforce the data smoothness in Z-axis. We set the total epoch to 5000. The total training time is 5 hours. 

\section{Results}

\subsection{Simulated experiment}
To validate the classification accuracy of the SRNN, we generated a simulated dataset with 5 different respiratory states sampling. For a real in-utero dataset we classified slices into motion states using combination of peak selection and manual input. We then reconstructed the average motion state and registered it to the acquired slices of the other respiratory states. A breathing cycle was then simulated based on the choice of eight slices from each group with random starting state. We tested the peak selection method and the SRNN to the simulated dataset. 

\begin{table}[]
\centering
\caption{Comparison respiratory state classification accuracy on the simulation dataset}
\begin{tabular}{p{2cm}|p{1.4cm}|p{1.4cm}|p{1.4cm}|p{1.4cm}|p{1.4cm}|p{2.5cm}}
\hline

Data ID & 1 & 2 & 3 & 4 & 5 & $Mean \pm Stdev $\\ \hline
Peak analysis & $39.61\%$ &$ 41.03\% $& $27.86\% $& $69.30\%$ & $47.72\% $& $45.10\% \pm 13.69\% $ \\ \hline
SRNN & $77.81\%$ & $77.00\%$ & $77.50\%$ & $77.60\%$ &$ 78.62\% $&$ 77.71\% \pm 0.53\% $\\ \hline
\end{tabular}
\label{tab:one}
\end{table}

Table 1 shows that SRNN achieved close to 80\% accuracy for all five breathing states, while original peak selection had much lower accuracy. This was mainly due to confusion of the neighbouring classes or average inhale and exhale states.

\subsection{Real data reconstructions}
MRI data were acquired on a 3T clinical system (Achieva, Philips Healthcare, Best, Netherlands) using a 2D bSSFP sequence with the SWEEP technique~\cite{LaurenceHJacksonAnthonyNPriceJanaHutterLucilioCordero-GrandeAlisonHoPaddyJSlator2018}. Informed consent was obtained from 2 healthy adult volunteers (kidney) and 10 pregnant volunteers (gestational ages: 23-36 weeks) who were scanned in the supine position with routine blood pressure and pulse oximetry monitoring. For kidney and uterus acquisitions, the TR/TE is 5.7/2.8 and 7.3/3.6 ms, the sweep rate is 0.37 and 0.17mm/s, and the slice thickness is 3 and 4mm, respectively.

The reconstruction results of the abdominal scan is shown in Fig. 2. For single orientation acquisition motion artifacts are present in SVR reconstruction in spite of the automatic rejection of misaligned slices (b, e). On the other hand, SR reconstruction of slices (c, f) selected using our proposed method resolved most of the breathing artifacts. The bottom row demonstrate the proposed SRNN can accurately separate the inhale and exhale respiratory states.
\label{sec:pagestyle}
\begin{figure}[t]
\addtolength{\tabcolsep}{+1pt} 
\includegraphics[width=0.95\textwidth]{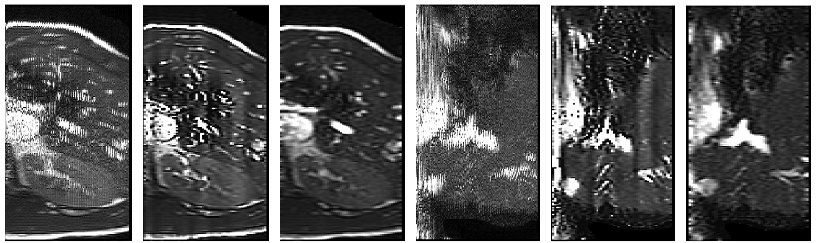}\\
\includegraphics[width=0.95\textwidth]{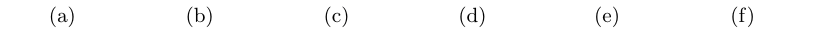}
\begin{tabular}{cccc}
\includegraphics[width=0.23\textwidth]{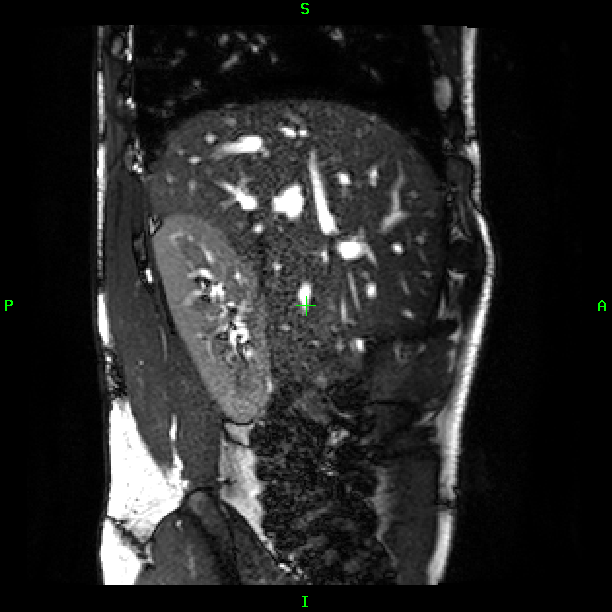}& 
\includegraphics[width=0.23\textwidth]{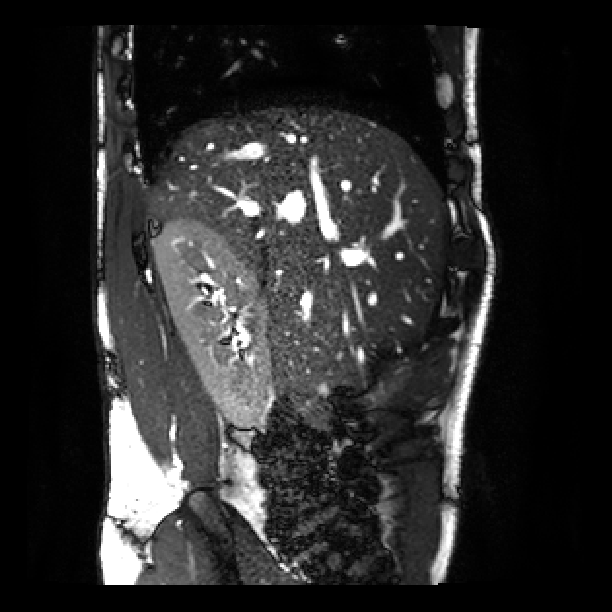}& 
\includegraphics[width=0.23\textwidth]{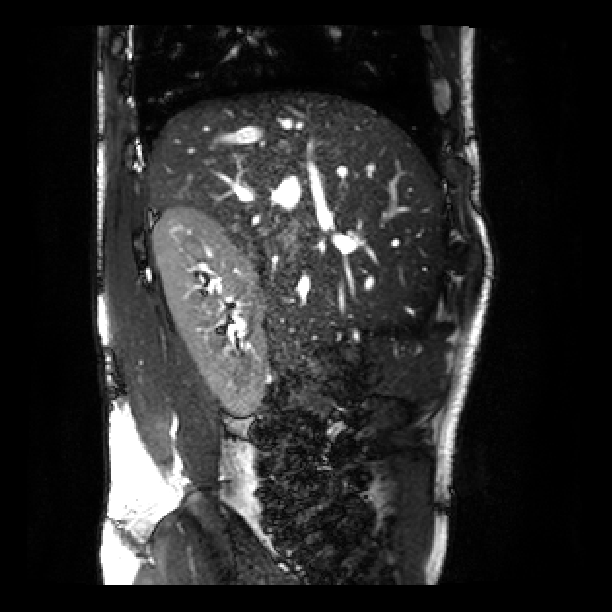}&
\includegraphics[width=0.23\textwidth]{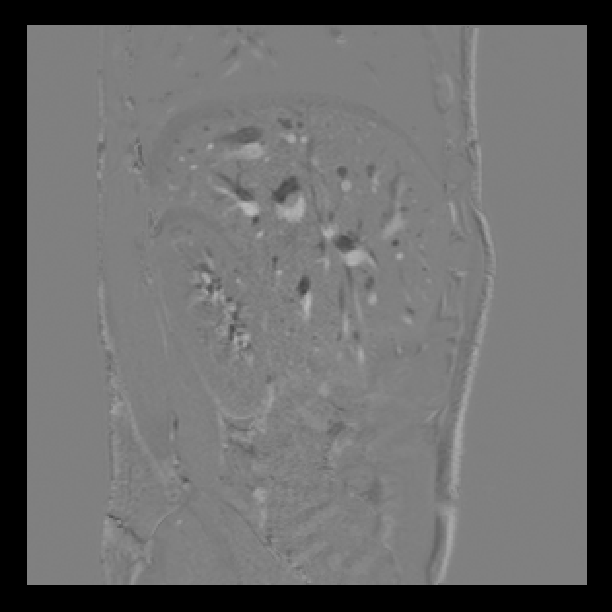}\\
\multicolumn{1}{c}{(h)}&{(i)}&{(j)}&{(k)}\\
\end{tabular}
\caption{Comparison reconstruction results of an abdominal subject. The top row shows the through-plane views:  the original motion corrupted MRI scan (a), (d); the SVR reconstruction (b), (e); and the propose reconstruction  (c), (f). The bottom row shows the Z-axis view of the original scan (h); the reconstruction results of inhale (i) and exhale state (j) and their difference (k)  }
\label{fig:one}
\end{figure}

Fig. 3 shows a similar comparison for abdominal MRI of a pregnant patient. As highlighted in the red box, where the artifact is caused by a deep breath, due to lack of a good target with only one stack of 2D MR images, the state-of-art SVR method failed in the area with large motion corruption.
\begin{figure*}
\begin{tabular}{ccc}
\includegraphics[width=0.321\textwidth]{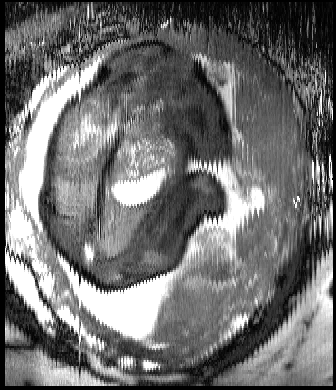}& 
\includegraphics[width=0.314\textwidth]{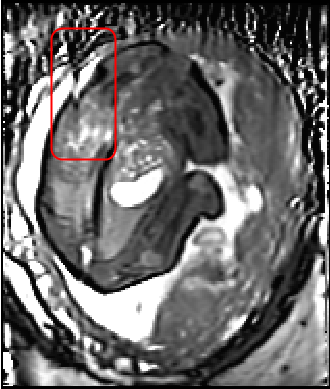}&
\includegraphics[width=0.318\textwidth]{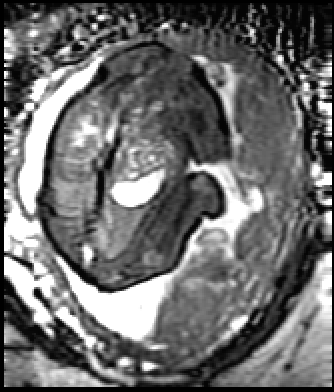}\\
\multicolumn{1}{c}{(a)}&{(b)}&{(c)}\\
\end{tabular}
\caption{Comparisons of perinatal subject with through-plane views: (a) the original motion corrupted image; (b) SVR based reconstruction; (c) the proposed reconstruction.}
\label{fig:two}
\end{figure*}

For quantitative analysis, we calculated the PSNR and SSIM between the reconstruction and sparsely selected 2D images with three different respiratory groups including average, inhale and exhale breathing states. We compared the proposed method with two state-of-the-art SVR software, the  SVR~\cite{Kuklisova-Murgasova2012} and NiftyMic~\cite{Ebner2018}. For fair comparison, we use the selected 2D images as a target and use free-form deformation (FFD) based method in MIRTK package.\footnote{https://mirtk.github.io/} to register the reconstructions from SVR and NiftyMic to each respiratory group. Our 4D reconstruction results are listed as SRNN0. We then register the average state in SRNN0 to the inhale and exhale states and report the results as SRNN1 in Table 2 and Table 3. The reported values in Table 2 and Table 3 are the average results of 10 in-utero subjects. The results show that the proposed reconstruction pipeline can generate SR images with high fidelity to the original MRI scan.

\begin{table}[]
    \centering
    \caption{Comparison PSNR results between the reference volume and the proposed reconstruction pipeline with different respiratory states}
    \begin{tabular}{|p{2cm}|p{2cm}|p{2cm}|p{2cm}|p{3cm}|}
    \hline
       & Average & Inhale  & Exhale & $Mean \pm  Stdev$\\
    \hline
        SVR~\cite{Kuklisova-Murgasova2012}& 32.82&32.48&31.59&	  $32.30\pm 0.64$
\\
 \hline
        NiftyMic~\cite{Ebner2018}& 31.43&32.29&30.98&$	31.57  \pm 0.67$
\\
\hline
        SRNN0& 36.87&36.95&35.28& $36.37 \pm 0.94$
\\
    \hline
        SRNN1 & - &	36.88	&35.28 &$	36.08 \pm 1.13$
\\
    \hline
    \end{tabular}
    
    \label{tab:two}
\end{table}
\begin{table}[]
    \centering
    \caption{Comparison SSIM results between the reference volume and the proposed reconstruction pipeline with different respiratory states }
  \begin{tabular}{|p{2cm}|p{2cm}|p{2cm}|p{2cm}|p{3cm}|}
    \hline
       & Average & Inhale  & Exhale & $Mean \pm  Stdev$\\
    \hline
        SVR~\cite{Kuklisova-Murgasova2012}& 0.94	&0.93&	0.93&	 $ 0.93\pm 0.01$
\\
 \hline
        NiftyMic~\cite{Ebner2018}& 0.92	&0.93&	0.91&$	0.92  \pm 0.01$
\\
\hline
        SRNN0& 0.96	&0.96 &	0.96&$ 0.96 \pm 0.01$
\\
    \hline
        SRNN1 &  -	&	0.96	&0.96&	$0.96 \pm 0.01$
\\
    \hline
    \end{tabular}
    
    \label{tab:three}
\end{table}

\section{Discussion and Conclusion}

In this paper we proposed an efficient respiratory motion resolved 4D (3D+t) reconstruction pipeline for abdominal and in-utero MRI. We investigated the respiratory information naturally embedded in the neighborhood slices and use it to train an bidirectional RNN. 

We propose a simple but effective motion correction and SR reconstruction pipeline for abdominal and in-utero MRI. The proposed pipeline can accurately cluster the respiratory motion of the acquired 2D images stack. The proposed self-supervised RNN utilise the NCC scores between each 2D slice and Z-axis blurred image. Such breathing motion indicator is very helpful to supervise the respiratory state clustering. The SR reconstruction stage further improves the reconstruction performances. Compared to SVRs, SRnet is a CNN pipeline that takes less than 1 min for a 4D reconstruction, while the SVR ones take around 40 mins \cite{Kainz2015} and 5 hours ~\cite{Ebner2018}. The PSNR and SSIM comparison results show that with such single orientation acquisition scenarios, the proposed pipeline with less than 20\% of the sparsely selected slices outperformed the SVR methods with all the slices. 
\section*{Acknowledgement}
This work was supported by the National Institutes of Health Human Placenta Project[1U01HD087202‐01], by the Wellcome Trust IEH Award [102431], by the Wellcome/EPSRC Centre for Medical Engineering [WT203148/Z/16/Z] and by the National Institute for Health Research (NIHR) Biomedical Research Centre at Guy’s and St Thomas’ NHS Foundation Trust and King’s College London. The authors also thank Nvdia for the GPU grant.
\bibliographystyle{splncs}

\bibliography{main}

\end{document}